\documentclass[12pt]{article}

\begin{document}
\renewcommand{\theequation}{\thesection.\arabic{equation}}
\title{Gravity on de-Sitter $3$-Brane, Induced Einstein-Hilbert Term and Massless Gravitons}
\author{J. Lukierski\footnote{Supproted by KBN grant 1P03B 01828}\, and I. Pr\'{o}chnicka \\ \\
Institute for Theoretical Physics, \\ University of Wroc\l{}aw, \\
 pl. M. Borna 9, 50-204 Wroc{\l}aw, Poland}

\date{}
\maketitle

\begin{abstract}
We study the extensions of DGP model which are described by five-dimensional Einstein gravity coupled covariantly to
$3$-brane with induced gravity term and consider warped D=4 de Sitter background field solutions on the brane. The case with included D=5 AdS cosmological term is also considered. Following background field method we obtain the field equations described by the Lagrangean
terms bilinear in gravitational field.
In such a linear field
approximation on curved dS background we calculate explicitly the five-dimensional massive terms as well as the mass-like ones on the brane.
We investigate the eigenvalue
 problem of Schr\"{o}dinger-like equation in fifth dimension for graviton masses and discuss the existence of massless as well as massive graviton modes in the bulk and on the brane without and with induced gravity.

\end{abstract}

\section{Introduction}

Following quite old ideas (see e.g. \cite{lukiz0} - \cite{lukiz2}), recently there was considered (see e.g \cite{lukiz3}--\cite{lukiz8})
four-dimensional spacetime as described by the $3$-brane
 moving
 in $D=5$ bulk. For realistic applications it is essential
 to study the properties of gravitational field on the brane, in
 particular, to investigate its mass spectrum. In a brane world scenario
 the $D=4$ gravity theory is replaced by multigravity \cite{lukiz24} with infinite set
 of new tensor fields describing the Kaluza-Klein (KK) modes in fifth dimension.
 An important question is the lowest mass value of the  multigravity
 quanta corresponding to the physical $D=4$ gravitons, in particular zero-mass graviton excitations. The appearance of nonvanishing graviton mass is linked with some important issues:

\begin{itemize}
\item Zakharov \cite{lukiz9} and van Dam, Veltman \cite{lukiz10} showed that in linearized gravity there is a mass discontinuity at graviton mass equal zero in 4D flat spacetime, what suggests that the astrophysical observations exclude massive gravity even for arbitrary small grawiton mass. However, recently it has been argued (see e.g.\cite{lukiz11}-\cite{lukiz12}) that one can restore the continuity of massive gravity theory at graviton mass $m=0$ by considering a curved maximally symmetric 4D spacetime (dS or AdS) with $\frac{m}{H}\rightarrow 0$ ($m$--mass parameter, $H$--Hubble parameter); the other way is to consider nonlinear effects, firstly pointed out by Vainstein \cite{lukiz13}, \cite{lukiz14}.

\item Since work by Boulware and Deser \cite{lukiz15} it is known that bilinear terms in gravitational field without derivatives which occur in the gravitational action can be decomposed into the one describing the cosmological term and the Pauli-Fierz graviton mass term. Pauli-Fierz term is the only one Lorentz-invariant mass-like term which after quantization does not generate ghosts in flat space. The cosmological term implies the readjustment of the Minkowski flat background to fixed curvature (dS or AdS) background (\cite{lukiz15}-\cite{lukiz18}) but the "curved" graviton remains massless, i.e. in four dimensions it has only two polarization states.
\end{itemize}

 We recall that the cosmological observations favour the choice of D=4 dS space as describing our universe (see e.g.\cite{lukiz38}). Subsequently, we
choose the warped  $D=5$ background with dS metric on the 3-brane and look for the mass eigenvalues of the `physical' graviton fluctuations. In Sect.2 we consider the extension of the DGP model \cite{lukiz5} with induced Einstein gravity term by adding the standard 3-brane action with nonvanishing tension. The considered model contains three constants: five-dimensional and four-dimensional Planck masses ($\kappa_5$, $\kappa_4$ respectively) and the brane tension parameter $T_3$. Furthermore, in order to find a link with RS model (\cite{lukiz4},\cite{lukiz37}) we add  in Sect.5 the five dimensional AdS cosmological term by introducing the fourth constant $\Lambda_5$.

Our main aim in this paper is to reproduce in an explicit way the derivation of massless and massive graviton modes. We show that in all these models\footnote{The only exception are the wave functions for the massive modes in the case $\Lambda_5\ne0$, $\kappa_5\ne0$, $\kappa_4\ne0$, $T_3\ne0$, which we have not found explicitly.} do exist the solutions describing massless and massive modes in the bulk. On the brane the field equations contain some additional singular terms which require suitable matching conditions on the bulk's boundaries . It appears that if we consider dS brane, one obtains the solutions of the equations valid also on the brane, however, the zero graviton modes are normalizable only if the induced gravity term vanishes ($\kappa_4\rightarrow\infty$).

 \section{DGP model with nonzero tension: warped de-Sitter background metric}
\setcounter{equation}{0}

 We investigate $D=5$ Einstein gravity with a vanishing bulk
 cosmological constant coupled convariantly to the positive tension ($T_3>0$) 3-brane (see
 e.g. \cite{lukiz19}--\cite{lukiz21}). With the inclusion of the brane tension we obtain the following generalisation of DGP model \cite{lukiz5}:

 \begin{eqnarray}\label{luiz1}
S &=& S^E+ S^E_{(ind)}+ S^{brane}=\frac{1}{\kappa_5^2} \int d^5\, \tilde{y}\, \sqrt{|G|} \, {\cal{R}}
\cr\cr &&   +
\frac{1}{\kappa_4^2} \int d^5\, \tilde{y} \, \delta(\tilde{z}) \, \sqrt{|g|} \, R_{(ind)}
\cr\cr && - \
   2 T_3 \int d^4 x \, \left\{
\sqrt{|g|} \, \left( g^{\mu\nu} (x) \partial_\mu \, \tilde{y}^A \, \,
\partial_\nu \, \tilde{y}^B\, G_{AB} (\tilde{y}) + 2
 \right)
    \right\}\, ,
\end{eqnarray}
where $A,B = 0,1,2,3,5$, $\mu,\nu=0,1,2,3$,
and
 $\tilde{y}^A (x)= (x^\mu, \tilde{z}(x))$ describes the position of $3$-brane in
 $5$-dimensional bulk space-time. We recall that the dimensionalities of $\kappa$ and $T_3$ are $[\kappa^2]=M^{-3}$ and  $[T_3]=M^4$. The induced metric
 $g_{\mu\nu}(x)$ on the $3$-brane is given in such a gauge by the
 formula:
 \begin{eqnarray}\label{luiz2}
g_{\mu\nu}(x) &= & \partial_\mu  \tilde{y}^A (x) \,  \partial_\nu  \tilde{y}^B (x)
\,
 G_{AB}(x_\mu , \tilde{z}(x))
 \cr\cr
 & = &
 G_{\mu\nu} (x_\mu , \tilde{z}(x))
  + 2 \partial_\mu \tilde{z}(x)
 G_{\mu 5}(x_\mu , \tilde{z}(x))
\cr\cr &&
 +
 \partial_\mu \, \tilde{z}(x) \, \partial_\nu \, \tilde{z}(x)  G_{55} (x_\mu , \tilde{z}(x))\, .
\end{eqnarray}

If we perform the transverse coordinate transformation $\tilde{z} \to z =
\tilde{z} -\tilde{z}(x)$, depending on the brane degrees of freedom, we obtain
the description of $3$-brane in a static gauge, i.e. straight and located in $z=0$ (or $y_A
=(x_\mu, 0)$), and from (\ref{luiz2}) follows that $g_{\mu\nu}(x) =
G_{\mu\nu}(x,0)$.

The field equations following from the action (\ref{luiz1}) in the static
gauge take the form (see e.g. \cite{lukiz12}):

i) the $D=5$ Einstein equations in the presence of $3$-brane
\begin{equation}\label{luiz3}
    E_{AB}\equiv {\cal{R}}_{AB}
    - \frac{1}{2}  \, G_{AB}{\cal{R}}+\lambda\,\delta(z)\left(R_{(ind)\mu\nu}
    - \frac{1}{2}  \, g_{\mu\nu}R_{(ind)}\right)\delta^\mu_A\delta^\nu_B= \frac{\kappa_5^2}{2} \, T_{AB}\,
\end{equation}
where $\lambda=\frac{\kappa_5^2}{\kappa_4^2}$, $\delta^\mu_{\ A}=\partial^\mu  y_A $, the energy-momentum tensor is given as:
\begin{equation}\label{luiz4}
T_{AB} = -2T_3 \sqrt{|g|}\, g_{\mu\nu}  \,
\delta^\mu_{A} \,  \delta^\nu_{B} \cdot \delta(z)\, .
\end{equation}
and taking the limit $\lambda \rightarrow 0$ means that induced gravity (IG) term vanishes.

It can be seen from equations (\ref{luiz3}) and (\ref{luiz4}) that in $D=4$ spacetime the background solution $g_{\mu\nu}$ can not be identified with a flat Minkowski metric $\eta_{\mu\nu}= diag(-,+,+,+)$.

ii) the equations of motion for $3$-brane
\begin{equation}\label{luiz5}
    \partial_\mu
    \left(
\sqrt{|g|}\, g^{\mu \nu}(x) \, g_{A \nu}(x)\right)
 - \frac{1}{2}
\sqrt{|g|}\  g_{\mu \nu}(x) \, \partial_A\,  g^{\mu\nu}(x) = 0
    \, ,
\end{equation}
where $g_{\mu 5}(x)= G_{\mu 5}(x,0)$ and
  $\partial_5 g_{\mu\nu} = \partial_z
 G_{\mu\nu}(x,z)\big|_{z=0}$.

Our aim here is to study the linear approximation of the field equation for
$E_{AB}(x)$ as following from (\ref{luiz3}) or quadratic
approximation to the action (\ref{luiz1}). The quadratic
approximation for the Einstein term is well-known: it gives the
free Lagrangean for $D=5$ massless graviton field. Our main point
is to take into consideration the new quadratic term coming from
the brane action.

In order to write down the background solution we
 use the ansatz for the line element in the bulk:
 \begin{eqnarray}\label{luiz6}
    ds^2 &=& g_{AB}^{(0)}(y) \, dy^A\, dy^{B} =
    g_{\mu\nu}^{(0)}(y) \, dx^{\mu}\, dx^{\nu} + dz^2 =
    \cr\cr
    &=&
    a^2 (z) \bar{g}^{(0)}_{\mu\nu} (x)\, dx^\mu \, dx^\nu + dz^2 \, ,
\end{eqnarray}
where $a(z)$ is a dimensionless warp factor.

The background metric (\ref{luiz6}) satisfies the equation:
\begin{equation}\label{luiz7}
R_{AB}^{(0)}-\frac{1}{2}g_{AB}^{(0)}R^{(0)}+\lambda\,\delta(z)\left(R^{(0)}_{(ind)\mu\nu}
    - \frac{1}{2}  \, g^{(0)}_{\mu\nu}R^{(0)}_{(ind)}\right)\delta^\mu_A\delta^\nu_B=\frac{\kappa^2}{2}T_{AB}^{(0)}\:,
\end{equation}
where $T_{AB}^{(0)}=T_{AB}|_{g_{CD}=g_{CD}^{(0)}}$.

We assume the following general background metric:
\begin{equation}\label{luiz9}
g_{AB}^{(0)}dy^A dy^B =a^2\left(z\right)\left(\eta_{00}dx^0 dx^0+ b^2\left(x_0\right)\eta_{ij}dx^i dx^i\right)+ dz^2 \:,
\end{equation}

If we put $T_3=0$ in eq. (\ref{luiz7}) (DGP model \cite{lukiz5}), the solution is provided by the flat Minkowski metric (a=b=1). In order to obtain the solution for $T_3\neq0$ we calculate the Riemann tensor $R^{(0)A}_{BPQ}$:
\begin{equation}\label{luiz10}
R^{(0)A}_{BPQ}=-\left(\delta ^{A}_0 R^{(0)1}_{BPQ} + \delta ^{A}_k R^{(0)2}_{BPQ}  + \delta ^{A}_5 R^{(0)3}_{BPQ}\right) \;,
\end{equation}
where
\begin{eqnarray}
R^{(0)1}_{BPQ} &=& \frac{a''}{a}\left[ \delta_{B}^5\delta_{P}^5\delta_{Q}^0-\delta_{B}^5\delta_{Q}^5\delta_{P}^0\right]+
\cr\cr &&
    +
\frac{b,_{00}}{ab} g_{ij}\left[ \delta_{P}^0\delta_{B}^i\delta_{Q}^j-\delta_{Q}^0\delta_{B}^i\delta_{P}^j
\right]+
\cr\cr    &&
-\left(\frac{a'}{a}\right)^2 g_{\mu\nu}\left[ \delta_{P}^0\delta_{B}^\mu\delta_{Q}^\nu-\delta_{Q}^0\delta_{B}^\mu\delta_{P}^\nu\right] \:,\nonumber
\end{eqnarray}
\begin{eqnarray}
R^{(0)2}_{BPQ} &=& \frac{a''}{a}
\left[ \delta_{Q}^k\delta_{P}^5\delta_{B}^5-\delta_{B}^5\delta_{Q}^5\delta_{P}^k\right]+
\cr\cr &&
+\frac{b,_{00}}{b}
\left[ \delta_{P}^0\delta_{B}^0\delta_{Q}^k-\delta_{Q}^0\delta_{B}^0\delta_{P}^k\right]
+
\cr\cr &&
+ \left( \frac{b,_0}{a b}\right)^2 g_{ij}
\left[\delta_{P}^k\delta_{B}^i\delta_{Q}^j - \delta_{P}^j\delta_{B}^i\delta_{Q}^k\right]
+
\cr\cr &&
- \left(\frac{a'}{a}\right)^2 g_{\mu\nu}\left[\delta_{P}^k\delta_{B}^\mu\delta_{Q}^\nu - \delta_{P}^\nu\delta_{B}^\mu\delta_{Q}^k\right] \:,
\nonumber
\end{eqnarray}

\begin{eqnarray}\nonumber
R^{(0)3}_{BPQ} &=& -\frac{a''}{a} g_{\mu\nu}\left[ \delta_{P}^5\delta_{Q}^\nu\delta_{B}^\mu-\delta_{Q}^5\delta_{B}^\mu\delta_{P}^\nu\right] \:,
\end{eqnarray}
and the prime denotes a derivative with respect to $z$, whereas $b,_{0}\equiv\frac{\partial b}{\partial x_0}$.

For the metric (\ref{luiz6}) with arbitrary $a(z)$ and arbitrary $b(x_0)$ we get:
\begin{eqnarray}\label{luiz31a}
R^{(0)\mu\nu}&=& -\left(aa''+ 3\left(a'\right)^2-3\left(\frac{b,_{0}}{b}\right)^2\right) \bar{g}^{(0)\mu\nu}
+\cr\cr &&
+\left(\frac{b,_{00}}{b}-\left(\frac{b,_0}{b}\right)^2\right)\left(\bar{g}^{ij}\delta^\mu_i\delta^\nu_j+3\bar{g}^{00}\delta^\mu_0\delta^\nu_0\right) \:.
\end{eqnarray}
We assume that there is a $Z_2$-symmetry in fifth dimension $y_5=z$ and we postulate
 (see e.g. \cite{lukiz11}) that the four-dimensional background
metric describes $D=4$ de-Sitter space so it has a form:
\begin{eqnarray}\label{luiz11}
\bar{g}^{(0)}_{\mu\nu}\, dx^\mu\, dx^\nu  &=& - dx^2_0 + b^2(x_0)\, dr^2
\cr\cr &=&
- dx^2_0 + e^{\frac{2}{c}x_0}\, dr^2 \,
\end{eqnarray}
($r=(x_1, x_2, x_3)$).
For the metric (\ref{luiz6}) with arbitrary $a(z)$ but $b(x_0)$ given by the formula (\ref{luiz11}) we get:
\begin{equation}\label{luiz31}
R^{(0)\mu\nu}  = -\left(aa''+ 3\left(a'\right)^2-3\left(\frac{b,_{0}}{b}\right)^2\right) \bar{g}^{(0)\mu\nu} \:.
\end{equation}

The metric given by (\ref{luiz6}), (\ref{luiz11}) is the solution of Einstein equations, provided that:
\begin{equation}\label{luiz12}
    a(z) =  \frac{1}{c_\lambda}\, \left(
    c_\lambda - |z|\right)\, ,
\end{equation}
and
\begin{equation}\label{luiz13}
    c_\lambda = \frac{3}{T_3 \kappa_5^2}\left(1+\sqrt{1+\frac{\lambda}{3}\kappa_5^2 T_3}\right) \,\, , \, \Lambda_4=\frac{3}{c_\lambda^2} ,
\end{equation}
where $D=4$ dS radius
\begin{eqnarray}
r_{dS}=\frac{1}{\Lambda_4}= \frac{6}{T_3^2 \kappa_5^4}\left(1+\sqrt{1+\frac{\lambda}{3}\kappa_5^2 T_3}\right) + \frac{\lambda}{T_3 \kappa_5^2} \, .
\end{eqnarray}
We see from the above formula that the inclusion of Einstein induced gravity term on the brane ($\lambda\neq 0$) does not change the functional form of the warp factor, given for all values $\lambda\geq 0$ by the formula (\ref{luiz12}).

For a background with a warped factor described by (\ref{luiz12}) one obtains:
\begin{equation}\label{luiz14}
T^{(0)}_{AB}=-2 \, T_3\sqrt{|g^{(0)}|}g^{(0)}_{\mu\nu}\delta ^{\mu}_A\delta ^{\nu}_B\delta(z) \;.
\end{equation}
The corresponding formula for $R^{(0)}_{AB}$ is:
\begin{equation}\label{luiz15}
R_{AB}^{(0)}=\left(\kappa_5^2 T_3-3\lambda \left(\frac{b_{,0}}{ab}\right)^2\right)\left(\frac{4}{3}g^{(0)}_{AB}-g^{(0)}_{\mu\nu}\delta^\mu_A\delta^\nu_B\right)\delta(z) \:.
\end{equation}
Let us notice that the Riemann tensor for the chosen metric (\ref{luiz11}) is proportional to $\delta(z)$, as well as the Ricci tensor, and is given by the formula:
\begin{equation}\label{luiz16}
R^{(0)A}_{BPQ}=-\left(\delta ^{A}_\mu R^{(0)1}_{BPQ} + \delta ^{A5}R^{(0)3}_{BPQ}\right) \;,
\end{equation}
where
\begin{eqnarray}
R^{(0)1}_{BPQ} &=& \frac{a''}{a}\left[ \delta_{B}^5\delta_{P}^5\delta_{Q}^\mu-\delta_{B}^5\delta_{Q}^5\delta_{P}^\mu\right]=
\cr\cr
    &=&
\frac{-2\delta(z)}{c_\lambda}\left[ \delta_{B}^5\delta_{P}^5\delta_{Q}^\mu-\delta_{B}^5\delta_{Q}^5\delta_{P}^\mu\right]
 \:,\nonumber
\end{eqnarray}

\begin{eqnarray}\nonumber
R^{(0)3}_{BPQ} &=& -\frac{a''}{a}g_{\mu\nu}\left[ \delta_{B}^\mu\delta_{P}^5\delta_{Q}^\nu-\delta_{B}^\mu\delta_{Q}^5\delta_{P}^\nu\right]=
\cr\cr
    &=&
\frac{2\delta(z)}{c_\lambda}g_{\mu\nu}\left[\delta_{B}^\mu\delta_{P}^5\delta_{Q}^\nu-\delta_{B}^\mu\delta_{Q}^5\delta_{P}^\nu\right]
 \nonumber
\end{eqnarray}
and $a(z)$ is given by (\ref{luiz12}), so that we have\footnote{Let us also notice that:

$R^{(0)}_{55}=-4 \frac{a''}{a}g_{55} \:.$}

\begin{equation}\label{luiz32}
R^{(0)\mu\nu}  = -aa''\bar{g}^{(0)\mu\nu} = \frac{2}{c_\lambda}\delta(z)\:,
\end{equation}

\begin{equation}\label{luiz33}
R^{(0)}  = R^{(0)A}_A= R^{(0)}_{\mu\nu}g^{\mu\nu}+R^{(0)}_{55}g^{55}=\frac{16}{c_\lambda}\delta(z) \:.
\end{equation}

\section{D=5 quadratic action and gauge fixing conditions}

\setcounter{equation}{0}
Let us consider the fluctuations $h_{AB}(y)$ ([$h_{AB}]=M^{\frac{3}{2}}$) of $D=5$ metric:

\begin{equation}\label{luiz17}
    g_{AB}(x,z) =  g_{AB}^{(0)}(x,z) + \kappa \, h_{AB}(x,z)\, .
\end{equation}

 Now we can substitute (\ref{luiz17}) in (\ref{luiz3}) and consider terms
 linear in $h_{AB}$. Correspondingly, the action (\ref{luiz1})
 can be expanded around the background solution (\ref{luiz6}) as follows (see
e.g. \cite{lukiz13,lukiz14}):
\begin{eqnarray}\label{luiz18}
    S\left[ g_{AB}\right] &= & S \left[ g_{AB}^{(0)}\right]
    + \kappa
    \int \frac{\delta S}{\delta g_{AB}(y) }
    \Bigg|_{g_{AB} = g_{AB}^{(0)}} h^{AB}(y) \, d^5y
    \\ \cr
    && +
    \kappa^2
     \int \frac{\delta S}{\delta g_{AB}(y_1)\delta g_{CD}(y_2) }
\Bigg|_{g_{AB} = g_{AB}^{(0)}}
\cr\cr
&& \cdot
\     h^{AB} (y_1)\, h^{CD} (y_2)\,
     d^5 y_1 \, d^5 y_2 + O(\kappa^3)\, .
     \nonumber
\end{eqnarray}

The free field approximation for $h_{AB}$ is following:
\begin{eqnarray}\label{luiz19}
&&S^E_{(2)}+S^E_{(ind)(2)}=\kappa_5^2\int \frac{\delta S^E}{\delta g_{AB}(y_1)\delta g_{CD}(y_2)} \Bigg|_{g_{AB} = g_{AB}^{(0)}}
     h^{AB} (y_1)\, h^{CD} (y_2)\, d^5 y_1 \, d^5 y_2+
 \cr\cr
     && \qquad
 +\kappa_4^2\int \frac{\delta S^E_{(ind)}}{\delta g_{AB}(y_1)\delta g_{CD}(y_2)} \Bigg|_{g_{AB} = g_{AB}^{(0)}}
     h^{AB} (y_1)\, h^{CD} (y_2)\, d^5 y_1 \, d^5 y_2+
 \cr\cr
     && \qquad
     =  \int d^5 y\sqrt{|g^{(0)}|}
     \left[-\frac{1}{4} \nabla^{(0)}_R h_{AB} \,
\nabla^{(0)R} h^{AB}
 + \frac{1}{4} \nabla^{(0)}_R \, h \, \nabla^{(0)R} h
\right.
 \cr\cr
     && \qquad
- \frac{1}{2}\nabla^{(0)}_B \, h \, \nabla^{(0)}_A\, h^{AB}
 +
\frac{1}{2}\nabla^{(0)}_A \, h_{BR} \, \nabla^{(0)R} h^{AB}+
  \cr\cr
    && \qquad
+\frac{1}{2}R^{(0)} \left(
\frac{1}{2} h^2 -   h_{AB} h^{AB}
\right)
 \cr\cr
  &&\qquad
 +
\left.
R^{(0)AB } \left(
2 h^R_B h_{AR} -   h \, h_{AB}
\right)
    \right]\, +
 \cr\cr
    && \qquad
    + \int d^5 y\sqrt{|g^{(0)}|}\lambda\delta(z)  \left[-\frac{1}{4} \nabla^{(0)}_\alpha h_{\mu\nu} \,
\nabla^{(0)\alpha} h^{\mu\nu}
 + \frac{1}{4} \nabla^{(0)}_\alpha \, \bar{h} \, \nabla^{(0)\alpha} \bar{h}
\right.
 \cr\cr
     && \qquad
- \frac{1}{2}\nabla^{(0)}_\nu \, \bar{h} \, \nabla^{(0)}_\mu\, h^{\mu\nu}
 +
\frac{1}{2}\nabla^{(0)}_\mu \, h_{\alpha\nu} \, \nabla^{(0)\alpha} h^{\mu\nu}+
  \cr\cr
    && \qquad
+\frac{1}{2}R_{(ind)}^{(0)} \left(
\frac{1}{2} \bar{h}^2 -   h_{\mu\nu} h^{\mu\nu}
\right)
 \cr\cr
  &&\qquad
 +
\left.
R_{(ind)}^{(0)\mu\nu } \left(
2 h^\alpha_\nu h_{\mu\alpha} -   \bar{h} \, h_{\mu\nu}
\right)
    \right]\, ,
\end{eqnarray}
where $h=h_A^A$, $\bar{h}=h_\mu^\mu$.

Using the fact that:
\begin{eqnarray}\nonumber
\nabla^R h^{AB}\nabla_A h_{RB}\equiv \nabla_Ah^{AB}\nabla^R h_{RB}+h^{AB}h^{PQ}R_{ABPQ}-h^{AB}h_{BP}R^P_A
\end{eqnarray}

and, additionally, that one can express Ricci tensor and Ricci scalar in terms of energy momentum tensor, for the arbitrary $a(z)$ and $b(x_0)$ given as in the eq. (\ref{luiz11}) one gets:

\begin{eqnarray}\label{luiz20}
&&S^E_{(2)}= \int d^5 y\sqrt{|g^{(0)}|}
     \Bigg[
-\frac{1}{4} \nabla^{(0)}_R h_{AB} \,
\nabla^{(0)R} h^{AB}
 + \frac{1}{4} \nabla^{(0)}_R \, h \, \nabla^{(0)R} h
 \cr\cr
 && \qquad
 -\frac{1}{2}\nabla^{(0)}_B \, h \, \nabla^{(0)}_A\, h^{AB}
 +
 \frac{1}{2}\left( \nabla_Ah^{AB}\nabla^R h_{RB}+h^{AB}h^{PQ}R_{ABPQ}-h^{AB}h_{BP}R^P_A \right)+
 \cr\cr
 && \qquad
 +\left(\kappa_5^2T_3-3\lambda\left(\frac{b,_0}{ab}\right)^2\right)\delta(z) \left(-2 h_{\mu R}h^{\mu R}+\bar{h}h+\frac{4}{3}h_{AB}h^{AB}-\frac{2}{3}h^2\right)         \Bigg]+
 \cr\cr
 && \qquad
+\int d^5 y\sqrt{|g^{(0)}|}\lambda\delta(z)
     \Bigg[
-\frac{1}{4} \nabla^{(0)}_\alpha h_{\mu\nu} \,
\nabla^{(0)\alpha} h^{\mu\nu}
 + \frac{1}{4} \nabla^{(0)}_\alpha \, \bar{h} \, \nabla^{(0)\alpha} \bar{h}
 \cr\cr
 && \qquad
 -\frac{1}{2}\nabla^{(0)}_\beta \, \bar{h} \, \nabla^{(0)}_\alpha\, h^{\mu\nu}
 +
 \frac{1}{2}\left( \nabla_\alpha h^{\mu\nu}\nabla^\alpha h_{\alpha\nu}-h^{\mu\nu}h_{\nu\beta}R^\beta_{(ind)\mu} \right)+
 \cr\cr
 && \qquad
 -\frac{3}{2}\left(\frac{b,_0}{ab}\right)^2 h_{\mu\nu}h^{\mu\nu}      \Bigg]
\end{eqnarray}
where raising and lowering of spacetime indices is done by means of the background metric $g^{(0)}_{AB}$, and $\nabla^{(0)}_R$ describes the covariant de-Sitter background derivative.
The part of the action for the brane takes the form:
\begin{eqnarray}\label{luiz21}
&&S^{brane}_{(2)}=\kappa_5^2\int \frac{\delta \, S^{brane}}{\delta \, g_{AB}(x_1)
\delta \, g_{CD}(x_2)}
\Bigg|_{g_{AB} = g_{AB}^{(0)}}
 h^{AB}
(x_1) \, h^{CD} (x_2) \,  d^4x_1 \,  d^4x_2=
\cr\cr
&& \qquad
=  T_3\kappa_5^2 \int d^4 x \sqrt{|g^{(0)}|}\left(
h_{\mu\nu} \,h^{\mu\nu}  - \frac{1}{2}\bar{h}^2
\right)=
\cr\cr
&& \qquad
=  \frac{3}{c_\lambda}\left(2+\frac{\lambda}{c_\lambda}\right) \int d^5 x \sqrt{|g^{(0)}|}\left(
h_{\mu\nu} \,h^{\mu\nu}  - \frac{1}{2}\bar{h}^2
\right)\delta(z)\:,
\end{eqnarray}
where $\bar{h}=h_\mu^\mu=h_{\mu\nu}g^{(0)\mu\nu}$, so that the whole action is the sum:
\begin{eqnarray}\label{luiz22}
&&S_{(2)}= \, \, S^E_{(2)}+S^{brane}_{(2)}
\;.
\end{eqnarray}
Varying the above action we get the equations for the field $h_{AB}$:
\begin{eqnarray}\label{luiz23}
 &&-\frac{1}{2}\left(\nabla^{(0)}_R \nabla^{(0)R}h_{AB}-g^{(0)}_{AB}\nabla^{(0)}_R \nabla^{(0)R} h + \nabla^{(0)}_A \nabla^{(0)}_B h
+ g^{(0)}_{AB}\nabla^{(0)R}\nabla^{(0)S} h_{RS}\right)
 \cr\cr
 &&
+ \frac{1}{2}\left(\nabla^{(0)R}  \nabla^{(0)}_A h_{RB}+\nabla^{(0)R} \nabla^{(0)}_B h_{AR}\right)+
\cr\cr
 &&
\Bigg[(-\frac{1}{2}\left(\nabla^{(0)}_\alpha \nabla^{(0)\alpha}h_{\mu\nu}-g^{(0)}_{\mu\nu}\nabla^{(0)}_\alpha \nabla^{(0)\alpha} \bar{h} + \nabla^{(0)}_\mu \nabla^{(0)}_\nu \bar{h}
+ g^{(0)}_{\mu\nu}\nabla^{(0)\alpha}\nabla^{(0)\beta} h_{\alpha\beta}\right)
 \cr\cr
 &&
+ \frac{1}{2}\left(\nabla^{(0)\alpha}  \nabla^{(0)}_\mu h_{\alpha\nu}+\nabla^{(0)\alpha} \nabla^{(0)}_\nu h_{\mu\alpha}\right)\Bigg]\lambda\delta^\mu_A\delta^\nu_B\delta(z)=
 \cr\cr
 &&
=\frac{6}{c_\lambda}\delta(z) \left(-2(h_{\mu B}\delta ^\mu_A+h_{\nu A}\delta ^\nu_B)-\frac{4}{3}h g^{(0)}_{AB}+\bar{h}g^{(0)}_{AB}+
\right.
 \cr\cr
 &&
\left.
+h g^{(0)}_{\mu\nu}\delta ^\mu_A\delta ^\nu_B+ 2\left( h_{\mu\nu} -\frac{1}{2} \bar{h}g^{(0)}_{\mu\nu}\right)\delta ^\mu_A\delta ^\nu_B+\frac{8}{3}h_{AB}\right)+
 \cr\cr
 && -\frac{3}{c_\lambda^2}\bar{h}g^{(0)}_{\mu\nu}\delta^\mu_A\delta^\nu_B\delta(z) + 3\left(\frac{b,_0}{ab}\right)^2 h_{\mu\nu}\lambda \delta(z)\:.
\end{eqnarray}

The bulk equations are covariant under five-dimensional infinitesimal general coordinate transformations
\begin{eqnarray}\label{luiz25}
\tilde{h}_{AB}\left(x,z\right)=h_{AB}\left(x,z\right)-\left(\nabla^{(0)}_{A}\chi_B\left(x,z\right)+\nabla^{(0)}_{B}\chi_A\left(x,z\right)\right) \:,
\end{eqnarray}
explicitly given in the form:
\begin{eqnarray}
\tilde{h}_{\mu\nu}&=& h_{\mu\nu}-\left(\partial_\mu \chi_\nu+\partial_\nu \chi_\mu -2\chi_A\Gamma^A_{\mu\nu}\right) \: ,\\
\tilde{h}_{\mu 5}&=& h_{\mu 5}-\left(\partial_\mu \chi_5+\partial_5 \chi_\mu -2\chi_\nu\Gamma^\nu_{\mu 5}\right) \: ,\\
\tilde{h}_{55}&=& h_{55}-2\partial_5 \chi_5 \:,
\end{eqnarray}
where
\begin{eqnarray}
\Gamma^\mu_{\nu 5}= \frac{a'}{a} g^\mu_\nu \:\:,
\Gamma^5_{\mu\nu}= -\frac{a'}{a} g_{\mu\nu} \:,\\
\Gamma^i_{0 j}= \frac{b,_0}{b} g^i_j \:\:,
\Gamma^{0}_{i j}= \frac{b,_{0}}{a^2b} g_{ij} \:.
\end{eqnarray}

In addition, there should be satisfied the orbifold symmetry conditions:
\begin{eqnarray}
h_{\mu\nu}(x,-z) &=& h_{\mu\nu}(x,z) \:,\\
h_{55}(x,-z)&=& h_{55}(x,z) \:, \\
h_{\mu 5}(x,-z) &=& - h_{\mu 5}(x,z) \:.
\end{eqnarray}
In the case under consideration one gets:
\begin{eqnarray}
&&\tilde{h}_{\mu\nu}= h_{\mu\nu}-\left(\partial_\mu \chi_\nu+\partial_\nu \chi_\mu +2\chi_5\frac{a'}{a}g^{(0)}_{\mu\nu}\right)+
 \cr\cr
 &&
 +
2\frac{b,_{0}}{ba^2}\left[\chi_0 \, g_{\mu\nu}^{(0)}\delta^{\mu}_i\delta^{\nu}_j+\chi_i\left(g^i_\mu\delta_{0\nu}+g^i_\nu\delta_{\mu 0}\right)\right] \: ,\label{luiz23a}\\
&&\tilde{h}_{\mu 5}= h_{\mu 5}-\left(\partial_\mu \chi_5+\partial_5 \chi_\mu -2\chi_\mu\frac{a'}{a}\right) \: ,\label{luiz23b}\\
&&\tilde{h}_{55}= h_{55}-2\partial_5 \chi_5 \:.\label{luiz23c}
\end{eqnarray}
It is easy to check that the gauge functions $\chi_M (x,z)$ also must satisfy the symmetry conditions:
\begin{eqnarray}
\chi_\mu(x,-z)&=& \chi_\mu(x,z) \:,\\
\chi_5(x,-z)&=& -\chi_5(x,z)\:.
\end{eqnarray}

Let us find the appropriate gauge that could be imposed on the field $h_{55}(x,z)$. First of all we choose $\chi_5 (x,z)$ in such a way, that it fulfils the above conditions, i.e.:
\begin{eqnarray}
\chi_5 (x,z)= \frac{1}{4}\int_{-z}^{z}h_{55}(x,\bar{z})d\bar{z} \:.
\end{eqnarray}

Now we can gauge away the function $\chi_5(x,z)$ so that eq.(\ref{luiz23b}) changes to:
\begin{eqnarray}
\tilde{h}_{\mu 5}= h_{\mu 5}-\left(\partial_5 \chi_\mu -2\chi_\mu\frac{a'}{a}\right) \:
\end{eqnarray}
and the important fact is that the brane is still straight in this gauge.

At the moment one can consider if it is possible to put $h_{\mu5}(x,z)=0$ everywhere. This field vanishes if
\begin{eqnarray}
\partial_5\left(a^{-2}\chi_\mu(x,z)\right)=a^{-2}h_{\mu5}(x,z)
\end{eqnarray}
so that the gauge function $\chi_\mu(x,z)$ takes the form:
\begin{eqnarray}
\chi_\mu(x,z)= a^2(z)\int_{-\infty}^{\infty} a^{-2}(\bar{z})h_{\mu5}(x,\bar{z})d\bar{z} \:,
\end{eqnarray}
and one still has a residual gauge transformation for $\chi_\mu(x,z)$ given by:
\begin{eqnarray}
\partial_5\left(a^{-2}\chi_\mu(x,z)\right)=0 \:.
\end{eqnarray}
As we have only one brane and do not consider matter on the brane, it seems to be possible, without brane bending effect, to assume that $h_{55}=0$.
Eventually, one gets:
\begin{eqnarray}
h_{\mu5}&=& 0 \:,\label{war1} \\
h_{55} &=& 0\nonumber \:.
\end{eqnarray}

\section{Equations of motion}
\setcounter{equation}{0}

Let us stress that the linear equation for $h_{AB}$ corresponds to the third term on rhs of (\ref{luiz18}), i.e. bilinear in $h_{AB}$.
Because $g_{AB}^{(0)}$ is the solution of $D=5$ Einstein equations, the
linear term in (\ref{luiz18}) disappears. Using the subsidiary conditions (\ref{war1}) the equations of motion have the following form:
\begin{itemize}
\item $(\mu\nu)$-component:
\begin{eqnarray}\label{luiz39dod}
 &&-\frac{1}{2}\Bigg[\nabla^{(0)}_R \nabla^{(0)R}h_{\mu\nu}+\nabla^{(0)}_\mu \nabla^{(0)}_\nu \bar{h}-\nabla^{(0)}_{\mu} \nabla^{(0)\rho} h_{\rho\nu}-\nabla^{(0)}_{\nu} \nabla^{(0)\rho} h_{\rho\mu}+
 \cr\cr
 &&
-g^{(0)}_{\mu\nu}\left(\nabla^{(0)}_R \nabla^{(0)R} \bar{h} -\nabla^{(0)\rho} \nabla^{(0)\sigma} h_{\rho\sigma}\right)\Bigg]+
 \cr\cr
 &&
 -\frac{1}{2}\Bigg[\nabla^{(0)}_\rho \nabla^{(0)\rho}h_{\mu\nu}+\nabla^{(0)}_\mu \nabla^{(0)}_\nu \bar{h}-\nabla^{(0)}_{\mu} \nabla^{(0)\rho} h_{\rho\nu}-\nabla^{(0)}_{\nu} \nabla^{(0)\rho} h_{\rho\mu}+
 \cr\cr
 &&
-g^{(0)}_{\mu\nu}\left(\nabla^{(0)}_\rho \nabla^{(0)\rho} \bar{h} -\nabla^{(0)\rho} \nabla^{(0)\sigma} h_{\rho\sigma}\right)\Bigg]\lambda\delta(z)-\frac{3}{c_\lambda^2}\bar{h}g^{(0)}_{\mu\nu}\lambda\delta(z)+
 \cr\cr
 &&
-\frac{2}{c_\lambda}\left(h_{\mu\nu}-\frac{1}{2}\bar{h}g^{(0)}_{\mu\nu}\right)\delta(z) - 3\lambda\left(\frac{b,_0}{ab}\right)^2 h_{\mu\nu}\delta(z)=0 \:.
\end{eqnarray}

\item $(\mu 5)$-component:
\begin{eqnarray}\label{luiz40adod}
\nabla^{(0)}_\mu\nabla^{(0)}_5 \bar{h}-\nabla^{(0)}_5\nabla^{(0)\nu} h_{\mu\nu}=0 \:,
\end{eqnarray}

\item $(5 \nu)$-component:
\begin{eqnarray}\label{luiz40bdod}
\nabla^{(0)}_5\nabla^{(0)}_\nu \bar{h}-\nabla^{(0)}_5\nabla^{(0)\mu} h_{\mu\nu}=0 \:,
\end{eqnarray}

\item $(55)$-component:
\begin{eqnarray}\label{luiz41dod}
&&-\frac{1}{2}\left(\nabla^{(0)}_5\nabla^{(0)}_5 \bar{h}+g^{(0)}_{55}\left(\nabla^{(0)\alpha}\nabla^{(0)\beta} h_{\alpha\beta}-\nabla^{(0)}_R\nabla^{(0)R} \bar{h}\right)\right)
\cr\cr
 &&
-\frac{2}{c_\lambda}\delta(z)\bar{h}g_{55}=0 \:,
\end{eqnarray}

\end{itemize}
From eq.(\ref{luiz40bdod}) one gets the constraint:
\begin{eqnarray}
\nabla^{(0)}_\nu\bar{h}=\nabla^{(0)\mu} h_{\mu\nu} \:,
\end{eqnarray}
which one can put into eq. (\ref{luiz40adod}) obtaining:
\begin{eqnarray}
\nabla^{(0)}_\mu\partial_5\bar{h}=\partial_5\nabla^{(0)}_\mu\bar{h} \:.
\end{eqnarray}
The lhs of the above equation is equal to $\partial_\mu\partial_5\bar{h}-\frac{a'}{a}\partial_\mu\bar{h}$, so that to be equal to the right hand side there must be satisfied the following condition:
\begin{eqnarray}
\bar{h}=0=\nabla^{(0)\rho}h_{\rho\nu} \,
\end{eqnarray}
 and as a consequence the above equations of motion simplify to:
\begin{itemize}
\item $(\mu\nu)$-component:
\begin{eqnarray}\label{luiz39dod1}
 &&-\frac{1}{2}\left(\nabla^{(0)}_R \nabla^{(0)R}h_{\mu\nu}
+\nabla^{(0)}_{\alpha} \nabla^{(0)\alpha} h_{\mu\nu}\lambda\delta(z)\right)+
\cr\cr
 &&
-\frac{2}{c_\lambda} h_{\mu\nu}\delta(z) - 3\lambda\left(\frac{b,_0}{ab}\right)^2 h_{\mu\nu}\delta(z)=0 \:,
\end{eqnarray}
\item $(55)$-component is identically equal to zero.
\end{itemize}

In curved 4D dS background embedded in a bulk, gravity fulfils the equation:
\begin{eqnarray}\label{kwiecien1}
\left(\nabla^{(0)}_\alpha \nabla^{(0)\alpha}+2\left(\frac{b,_0}{ab}\right)^2\right)\bar{h}_{\mu\nu}(x) = \frac{m^2}{a^2} \bar{h}_{\mu\nu}(x) \,.
\end{eqnarray}
This shift corresponds to the 4D cosmological constant and is required for recovering the equation of motion for a massless graviton if only $m^2=0$. We immediately notice that one can make the following separation of variables:
\begin{eqnarray}\label{kwiecien3}
h_{\mu\nu}\left(x,z\right) = \bar{h}_{\mu\nu}\left(x\right)\cdot \phi_m(z) \,.
\end{eqnarray}
Using above formulae and the fact that for dS brane embeddend in Minkowski bulk $\left(\frac{a'}{a}\right)^2=\left(\frac{b,_0}{ab}\right)^2$, equation (\ref{luiz39dod1}) takes the form:
\begin{eqnarray}\label{kwiecien2}
\left(-\partial_5\partial^5 + 2\frac{a''}{a}+ 2\left(\frac{a'}{a}\right)^2 - 4\left(\frac{a'}{a}\right)^2 \lambda\delta(z)\right) \phi_m\left(z\right) = \frac{m^2}{a^2}\left(1+\lambda\delta(z)\right) \phi_m\left(z\right) \,\,.
\end{eqnarray}

Using the eq. (\ref{kwiecien2}) one can examine the spectrum of graviton modes, where the $m=0$ ground state at $z=0$ corresponds to the amplitude of the physical graviton on the brane. The graviton localisation on the brane effectively can be described by the normalisability of the wave function $\phi_0(z)$, i.e. the condition:
\begin{eqnarray}\label{znormalizacja}
\int dz\, a^{-2}(z)|\phi_0(z)|^2<\infty \,.
\end{eqnarray}

In order to see whether the solution of the eq. (\ref{kwiecien2}) is a normalizable function, it is useful to express this equation in terms of conformal variables, i.e.:
\begin{eqnarray}\label{kwiecien5a}
 w = \, \int\frac{dz}{a(z)} \, ,
\end{eqnarray}
\begin{eqnarray}\label{kwiecien5b}
 \bar{\phi}_m(w) = \left(a(z)\right)^{-\frac{1}{2}}\phi_m(z)\, ,
\end{eqnarray}
whereas
\begin{eqnarray}\label{kwiecien5c}
 a(w)=e^{-\sqrt{\frac{\Lambda_4}{3}} |w|} = e^{A(w)} \, .
\end{eqnarray}
We get explicitly
\begin{eqnarray}\label{kwiecien5d}
w = -\epsilon(z)\, c_\lambda \, \ln \frac{c_\lambda}{c_\lambda-|z|}
\end{eqnarray}
and
\begin{eqnarray}\label{kwiecien5e}
z = \epsilon(w) c_\lambda \left(1-e^{-\frac{1}{c_\lambda}|w|}\right)\,  \,,
\end{eqnarray}
where the integration constant in (\ref{kwiecien5a}) is arranged in such a way that for $z\rightarrow 0$ we get $w\rightarrow 0$.
As a result, one obtains Schr\"{o}dinger-like equation
\begin{eqnarray}\label{kwiecien6}
\left(-\frac{d^2}{dw^2} + V(w)\right) \bar{\phi}_m\left(w\right) = m^2\left(1-C_2\lambda\delta(w)\right) \bar{\phi}_m\left(w\right) \, ,
\end{eqnarray}
with the potential $V(w)$ given by the formula:
\begin{eqnarray}
V(w) = \frac{9}{4} \left(\frac{dA}{dw}\right)^2\left(1-C_1\lambda\delta(w)\right) + \frac{3}{2} \frac{d^2 A}{dw^2} \, ,
\end{eqnarray}
where $C_1$, $C_2$ are positive constants and the condition (\ref{znormalizacja}) is reexpressed in the following way:
\begin{eqnarray}\label{wnormalizacja}
\int dw\, |\phi_0(w)|^2<\infty \,.
\end{eqnarray}

\section{Graviton on the 3-brane without and with induced gravity}
\setcounter{equation}{0}

\subsection{dS$_4$/M$_5$ case without and with IG term}

  We consider two cases:

\begin{itemize}
\item  $\lambda=0$

In the limit eq. (\ref{kwiecien2}) takes the following form:
\begin{eqnarray}\label{li1}
\left(-\partial_5\partial^5 + 2\left(\frac{a'}{a}\right)^2 + 2\frac{a''}{a}\right) \phi_m\left(z\right) = \frac{m^2}{a^2} \phi_m\left(z\right) \,\,,
\end{eqnarray}
and $a(z) = \frac{1}{c}\left(c-|z|\right)$, $c = \lim_{\lambda\rightarrow0} c_\lambda$. Explicitly, we get:
\begin{eqnarray}\label{li1a}
\left(-\partial_5\partial^5 + \frac{2}{(c-|z|)^2} - \frac{4}{c}\delta(z)\right) \phi_m\left(z\right) = \frac{m^2 c^2}{(c-|z|)^2} \phi_m\left(z\right) \,\,.
\end{eqnarray}
The general solution of (\ref{li1}) with  the appropriate $a(z)$ is given as:

\begin{itemize}
\item zero mode ($m=0$)
\begin{eqnarray}\label{li2}
\phi_0\left(z\right) \sim a^2(z) \, .
\end{eqnarray}
Let us notice an important fact that the zero mode is expressed in terms of variable $z$ in a form, which is also correct for the 4D dS brane embedded in any background (see e.g.\cite{lukiz28}) (AdS$_5$ or dS$_5$) with correspondingly given function $a(z)$.

\item massive modes ($m>0$)
\begin{eqnarray}\label{li3}
\phi_m\left(z\right) \sim a^{\frac{1}{2}}(z)\left(a^{\frac{3}{2}i\sqrt{\frac{4}{9}m^2c^2-1}}(z) + K_m \, a^{-\frac{3}{2}i\sqrt{\frac{4}{9}m^2c^2-1}}(z)\right) \, ,
\end{eqnarray}
where
\begin{eqnarray}\label{li4}
K_m =  \frac{i\sqrt{\frac{4}{9}m^2c^2-1}-1}{i\sqrt{\frac{4}{9}m^2c^2-1}+1} \, .
\end{eqnarray}
The solution exists only for $m^2>\frac{9}{4c^2}$.
\end{itemize}

On the other hand, with the use of the variables (\ref{kwiecien5a})-(\ref{kwiecien5c}) when $\lambda\rightarrow 0$ (see also \cite{lukiz21})  one can see easily that the potential has the following explicit form:
\begin{eqnarray}
V(w) = \frac{3}{4} \Lambda_4 - 3 \sqrt{\frac{\Lambda_4}{3}}\delta(w) \,.
\end{eqnarray}
We observe that this potential has the same functional form as the flat space volcano potential, i.e. we can obtain the normalizable modes trapped on the brane \cite{lukiz27}.
The equation (\ref{kwiecien6}) has the following solutions:

\begin{itemize}
\item zero mode ($m=0$)

\begin{eqnarray}\label{wrzesien1}
\bar{\phi}_0\left(w\right) \sim e^{-\frac{3}{2c}|w|}\, ,
\end{eqnarray}

\item massive modes ($m\neq0$)

\begin{eqnarray}\label{wrzesien2}
\bar{\phi}_m\left(w\right) \sim e^{-\frac{3}{2c}i\sqrt{\frac{4}{9}m^2c^2-1}|w|}\,+ K_m \, e^{\frac{3}{2c}i\sqrt{\frac{4}{9}m^2c^2-1}|w|} \,,
\end{eqnarray}
where $K_m$ is given by eq.(\ref{li4})\footnote{It should be stressed here that the solution presented by Ito in \cite{lukiz21} has only the first term of (\ref{wrzesien2}), what does not produce the solution at $z=0$.}.
\end{itemize}

 In the dS-brane case the potential approaches a constant value when $w$ tends to infinity and there is a mass gap\footnote{At $\lambda=0$, $\Lambda_4 = \frac{3}{c^2}$.} $m^2>\frac{3}{4}\Lambda_4$, between the zero mode and massive modes of graviton in the continuum due to consistency with matching conditions. One can notice that this gap in the continuum is connected, by the constant value of the potential at infinity, with $\Lambda_4$, i.e. if the value of $\Lambda_4$ is smaller, the width of the gap becomes narrower.

\item  $\lambda\neq0$

The equation (\ref{kwiecien2}) has the following explicit form:
\begin{eqnarray}\label{li1b}
&& \left(-\partial_5\partial^5 + \frac{2}{(c_\lambda-|z|)^2} + \frac{4\lambda}{c_\lambda^2}\delta(z)- \frac{4}{c_\lambda}\delta(z)\right) \phi_m\left(z\right) =
\cr\cr &&
=\frac{m^2 c_\lambda^2}{(c_\lambda-|z|)^2}\left(1+ \lambda\delta(z)\right)\phi_m\left(z\right) \,\,.
\end{eqnarray}
It is easy to notice that in the considered case for $z\neq0$ one obtains the solutions of the above equation in the same functional form as in the case with $\lambda=0$, i.e. (\ref{wrzesien1}) for $m=0$ and (\ref{wrzesien2}) for $m^2>\frac{3}{4}\Lambda_4$, but with $c_\lambda$ (see eq.(\ref{luiz13})) replacing $c$. The question is if one can extend these solutions for the whole range of $z$, i.e. including the brane position at $z=0$.

\begin{itemize}
\item zero mode ($m=0$)

 We shall make the following general ansatz:
\begin{eqnarray}\label{ansatz}
\phi_0\left(|z|;\lambda\right) \sim a^2(|z|)f(|z|;\lambda)\, ,
\end{eqnarray}
\begin{eqnarray}\label{ansatzA}
f\left(|z|;0\right) = 1\, .
\end{eqnarray}
From eq. (\ref{kwiecien2}) taken for $m=0$, one gets two equations :
\begin{eqnarray}
4\dot{a}\dot{f} & = & -a\ddot{f} \label{1} \;, \\
a^2\dot{f}\delta(z) & = & -2(\dot{a})^2 f\lambda\delta(z) \label{2}\;,
\end{eqnarray}
where $\dot{a}\equiv \frac{\partial{a}}{\partial{u}}$ etc., $u\equiv |z|$.
The eq. (\ref{1}) gives the solution in the following form:
\begin{eqnarray}
f(|z|;\lambda)= - \, D\frac{c^4_\lambda}{3(c_\lambda-|z|)^3}+E \;
\end{eqnarray}
where $D$ is an arbitrary constant, whereas the eq. (\ref{2}) defines the integration constant $E$ as:
\begin{eqnarray}
E = -D \,c_\lambda \frac{2\lambda + 3c_\lambda}{6\lambda} \;.
\end{eqnarray}
Due to the condition (\ref{ansatzA}) one notices immediately that $E\rightarrow -\infty$, so we can not find the physical solution unless we assume that $D=-\frac{6\lambda}{2\lambda c_\lambda + 3c^2_\lambda}$. In this case $E=1$ and $f(|z|;\lambda)$ takes the form:
\begin{eqnarray}
f(|z|;\lambda) = 1-\frac{2\lambda}{2\lambda+3c_\lambda}a^{-3}(z)
\end{eqnarray}
satisfying the condition (\ref{ansatzA}). Subsequently, for arbitrary $z$ one gets:
\begin{eqnarray}\label{rozwiazanie0}
\phi_0(|z|;\lambda) \sim a^2(z)\left(1-\frac{2\lambda}{2\lambda+3c_\lambda}a^{-3}(z)\right)
\end{eqnarray}
but this solution is not normalizable what can be shown using the conformal variable $w$.\footnote{The solution (\ref{rozwiazanie0}) is expressed in a variable $w$ as:
\begin{eqnarray}
\bar{\phi}_0\left(w;\lambda\right) \sim e^{-\frac{3}{2c_\lambda}|w|} + \frac{2\lambda}{2\lambda + 3c_\lambda}e^{+\frac{3}{2c_\lambda}|w|} \,.\nonumber
\end{eqnarray}}

The lack of the normalizable graviton wave function can be demonstrated under plausible technical assumptions if we assume that normalizable solution in the presence of the induced term, i.e. $\lambda\neq 0$,  has the following form:
\begin{eqnarray}
\bar{\phi}_0\left(w;\lambda\right) \sim e^{-B(\lambda)|w|}\,
\end{eqnarray}
with the condition $B\left(0\right)=\frac{3}{2c}$. From the eq. (\ref{kwiecien6}) (for $m=0$) we obtain the following two equations:
\begin{eqnarray}
B & = & ^{+}_{-} \frac{3}{2c_\lambda} \, ,\\
0 & = & \left(2B -\frac{3}{c_\lambda} - \frac{9C_1}{4c_\lambda^2}\lambda\right)\delta(w)\,.\label{warunek}
\end{eqnarray}
The normalizability of the function $\bar{\phi}_0\left(w;\lambda\right)$ requires to choose positive $B$, but then, from eq. (\ref{warunek}), it follows immediately that $\lambda=0$.

\item massive modes ($m>0$)

In this case, for $z=0$ one gets the same functional form of the solution as for $z\neq 0$, i.e. (\ref{wrzesien2}) with the condition $m^2>\frac{3}{4}\Lambda_4$, but instead of the constant $K_m$ one should put explicitly:
\begin{eqnarray}
K_m (\lambda)= \frac{3i\sqrt{\frac{4}{9}m^2c^2_\lambda-1}-3-\lambda\left(4+m^2c_\lambda\right)}{3i\sqrt{\frac{4}{9}m^2c^2_\lambda-1}+3+\lambda m^2c_\lambda} \,.
\end{eqnarray}
In the limit $\lambda\rightarrow0$ $K_m (0)=K_m$ is given by the formula (\ref{li4}).
\end{itemize}

\end{itemize}

\subsection{dS$_4$/AdS$_5$ case without and with IG term}

We start with the action given by (\ref{luiz1}) in a static gauge with included 5-dim cosmological constant term:
\begin{eqnarray}\label{marzec1}
S &=& S^E+ S^E_{(ind)}+ S^{brane}=\frac{1}{\kappa_5^2} \int d^5\, x\, \sqrt{|g|} \, {\cal{R}}+
\cr\cr &&   -
\frac{2}{\kappa_5^2} \int d^5\, x\, \sqrt{|g|} \, \Lambda_5
\cr\cr &&   +
\frac{1}{\kappa_4^2} \int d^5\, x\, \delta(z) \, \sqrt{|g|} \, R_{(ind)}
\cr\cr && - \
   2 T_3 \int d^5 x \, \sqrt{|g|} \, \delta(z) \,.
\end{eqnarray}

The Einstein equations (see (\ref{luiz3})-(\ref{luiz4})) are extended as follows:
\begin{equation}\label{marzec2}
    {\cal{R}}_{AB}
    - \frac{1}{2}  \, G_{AB}\left({\cal{R}}-\Lambda_5\right)+\lambda\,\delta(z)\left(R_{(ind)\mu\nu}
    - \frac{1}{2}  \, g_{\mu\nu}R_{(ind)}+\kappa_4^2T_3g_{\mu\nu}\right)\delta^\mu_A\delta^\nu_B= 0\,,
\end{equation}
where $\lambda \equiv \frac{\kappa_5^2}{\kappa_4^2}$. As a background metric we choose:
\begin{eqnarray}\label{marzec3}
ds^2 = a^2(z)\bar{g}_{\mu\nu}^{(0)}(x)dx^{\mu}dx^{\nu} + dz^2 \, ,
\end{eqnarray}
where
\begin{eqnarray}\label{marzec4}
\bar{g}_{\mu\nu}^{(0)}(x)dx^{\mu}dx^{\nu} &=& -dx_0^2 + b^2(x_0) dr^2 =
\cr\cr &=&
-dx_0^2 + e^{\frac{2}{c}x_0} dr^2,
\end{eqnarray}
describes dS metric on the brane and
\begin{eqnarray}\label{marzec5}
a(z) = \frac{\sinh(A(z_0-|z|))}{\sinh(Az_0)} \, ,\,\,\; \, A =\sqrt{-\frac{\Lambda_5}{12}}\,
\end{eqnarray}
where $z_0$ is a positive constant.
The four parameters $\kappa_5$, $\kappa_4$, $T_3$ and $\Lambda_5$ which appear in eq. (\ref{marzec1}) are linked by the relation:
\begin{eqnarray}\label{marzec6}
T_3 = \frac{6}{\kappa_5}\sqrt{\Lambda_4 - \frac{\Lambda_5}{12}} \,  + \frac{\Lambda_4}{\kappa_4^2}\, .
\end{eqnarray}
 Taking the limit $\Lambda_4\rightarrow 0$ of the eq. (\ref{marzec6}) one obtains the Randall-Sundrum tunning condition in RS model. If $\Lambda_4 > 0$ we have dS$_4$ 3-brane embedded in AdS$_5$ spacetime.

For the background with the metric (\ref{marzec4})-(\ref{marzec5}) one obtains:
\begin{eqnarray}\label{marzec8}
{\cal{R}}_{AB}^{(0)} = \left(\kappa_5^2T_3 - 3\lambda\left(\frac{b,_0}{b}\right)^2\right)\left(\frac{4}{3}g_{AB}^{(0)} - g_{\mu\nu}^{(0)}\delta^\mu_A\delta^\nu_B\right)\delta(z) + \frac{\Lambda_5}{3}g_{AB}^{(0)}\,,
\end{eqnarray}

\begin{eqnarray}\label{marzec9}
{\cal{R}}^{(0)} = \frac{8}{3}\left(\kappa_5^2T_3 - 3\lambda\left(\frac{b,_0}{b}\right)^2\right)\delta(z) + \frac{5\Lambda_5}{3} \,.
\end{eqnarray}
Following the calculations from Section 3 and 4 in the same gauge we get the equations for the fluctuations $h_{\mu\nu}(x,z)$, i.e.:
\begin{eqnarray}\label{marzec10}
 &&\nabla^{(0)}_R \nabla^{(0)R} \,h_{\mu\nu}
+\nabla^{(0)}_{\alpha} \nabla^{(0)\alpha}\, h_{\mu\nu}\,\lambda\,\delta(z)+
\cr\cr
 &&
+\frac{2}{3} \left(\kappa_5^2 T_3 - 3\lambda\left(\frac{b,_0}{ab}\right)^2\right) h_{\mu\nu}\delta(z) + \frac{1}{3}\Lambda_5  h_{\mu\nu}=0 \:.
\end{eqnarray}
Using Einstein equations for $g_{AB}^{(0)}$, i.e.:
\begin{itemize}
\item $(\mu\nu)$-component:
\begin{eqnarray}\label{marzec11}
\frac{a''}{a} + \left(\frac{a'}{a}\right)^2 - \left(\frac{b,_0}{ab}\right)^2 = -\frac{\Lambda_5}{6}-\left(\frac{\kappa_4^2}{3} \,T_3\, +  \left(\frac{b,_0}{ab}\right)^2\right)\lambda\delta(z) \, ,
\end{eqnarray}
\item $(55)$-component:
\begin{eqnarray}\label{marzec12}
\left(\frac{a'}{a}\right)^2 - \left(\frac{b,_0}{ab}\right)^2 = -\frac{\Lambda_5}{12} \, ,
\end{eqnarray}
\end{itemize}
and the equation for gravity in curved 4D background, given by the eq. (\ref{kwiecien1}), one can separate variables according to eq. (\ref{kwiecien3}) to get from eq. (\ref{marzec10}):

\begin{eqnarray}\label{marzec13}
&& \left(-\partial_5\partial^5 + 2\frac{a''}{a} + 2\left(\frac{a'}{a}\right)^2 -4 \left(\left(\frac{a'}{a}\right)^2 + \frac{\Lambda_5}{12}\right) \lambda\delta(z)\right)\,\phi_m(z) =
\cr\cr
 && = \frac{m^2}{a^2}\left(1+\lambda\delta(z)\right) \,\phi_m(z) \,.
\end{eqnarray}
We see that the eq. (\ref{marzec13}) differs from the eq. (\ref{kwiecien2}) only by a local term proportional to $\Lambda_5$, however, it should be pointed out that the warp factor $a(z)$ occuring in eq. (\ref{marzec13}) is given by $\Lambda_5$-dependent function (\ref{marzec5}).

One can calculate the following solutions for arbitrary $z$ (i.e. in the bulk  at $z\neq0$ and on the brane at $z=0$):

\begin{itemize}

\item $\lambda= 0$ (see also \cite{lukiz28})

\begin{itemize}

\item massless mode ($m=0$)

The formula (\ref{li2}), i.e. $\phi_0(z) \sim a^2(z)$, is also valid here, however, $a(z)$ is expressed by (\ref{marzec5}).

\item massive modes ($m>0$)

The formula (\ref{li3}) is modified as follows:
\begin{eqnarray} \label{hiper1}
 \phi_m(z) \sim a^{-2}(z)\,\, _{2}F_{1}\left(U_1 , \, U_2 ; \,3 ; \,U_3 \right) \,,
\end{eqnarray}
where
\begin{eqnarray}\label{hiper2}
U_1 & = & \frac{5}{4}\mp\frac{3i}{4} \sqrt{\frac{4}{9}m^2c^2-1}\label{hiper3} \,\,,\\
U_2 & = & \frac{5}{4}\pm\frac{3i}{4}\sqrt{\frac{4}{9}m^2c^2-1} \label{hiper4}\,\,,\\
U_3 & = & -a^{-2}(z)\sinh^{-2}(Az_0)  \label{hiper5}\,\,.
\end{eqnarray}
The solutions exist for $m^2>\frac{3}{4}\Lambda_4$.
\end{itemize}

\item $\lambda\neq0$

\begin{itemize}

\item massless mode ($m=0$)

As in the case of dS$_4$/M$_5 +$IG one can assume that:

\begin{eqnarray}
\phi_0(|z|;\lambda) \sim a^2(|z|)f(|z|;\lambda)
\end{eqnarray}
where we assume (\ref{ansatzA}) and take $a(|z|)$ given by (\ref{marzec5}). One can show that the eq. (\ref{marzec13}) with $m=0$ is satysfied by the following solution:

\begin{eqnarray}
\phi_0(|z|;\lambda) &\sim& \frac{\sinh^2(A(z_0-|z|))}{\sinh^2(Az_0)}\, (\, 1+
\cr \cr
&&
+\lambda \frac{4A^2\cosh^2(Az_0)+\frac{\Lambda_5}{3}}{6A}\sinh^4(Az_0)\tanh(A(z_0-|z|))
\cr \cr
&&
\left(\cosh^{-2}(A(z_0-|z|))+2\right)+
\cr \cr
&&
-\lambda\frac{1}{6}\sinh^5(Az_0)(\cosh^{-2}(A(z_0-|z|))+2)
\cr \cr
&&
\left(4A\cosh(Az_0)+\frac{\Lambda_5}{3A\cosh(Az_0)}\right)) \,.
\end{eqnarray}

\end{itemize}
The explicit form of the wave functions with $m>0$, generalising for $\lambda\neq 0$ the formulae (\ref{hiper1})-(\ref{hiper5}), is under consideration, but one can conjecture that it is also given by a hypergeometric function.
\end{itemize}

\section{Final remarks}

Since the work of Dvali, Gabadadze, Porrati \cite{lukiz5} brane-world models with induced gravity focused a lot of attention because they provide a reasonable modification of gravity at high energies. The main challenge facing the brane-world models appears to be the agreement with the current cosmological data, in particular with the accelerating expansion of the Universe. The other problem is the existence of explicit normalizable solution describing the massless graviton mode. The mass spectrum of gravitons in RS (fine-tuned) model with induced term was widely discussed in many papers (see f.e. \cite{lukiz28}-\cite{lukiz30}). In such a model the brane-localized curvature terms do not modify the equation for massless graviton mode, however they do change the eigenvalue equation for massive gravitons. Our aim was to investigate how this result is generalised for detuned models with induced gravity on dS-like brane in flat as well as AdS bulk (compare e.g. with \cite{lukiz31}). Our hopes about the existance of solutions were supported by general arguments recently presented by Kaloper (\cite{lukiz32}).

We list in this paper the results obtained for the de-Sitter 3-brane without and with induced Einstein term embedded in a $D=5$ Minkowski or AdS$_5$ bulk. We found that even for a massless graviton  mode the induced gravity present in the action modifies the equations of motion  in such a way that we could not find the wave functions with the appropriate physical normalization conditions. However, there were found new explicit solutions for massless and massive modes for $z\neq0$ with their extensions on the brane, i.e. satisfying the matching conditions at $z=0$. Unfortunately, the corresponding wave functions in the sense of relations (\ref{znormalizacja}) or (\ref{wnormalizacja}) are not normalizable.

We would like to add that we limited calculations in this paper to linear graviton field equations (quadratic action) and we did not study the possible appearence of ghosts for quantum fluctuations. An interesting problem is to study in the case of $\lambda\neq 0$ the higher orders of perturbations around the background solution (see e.g. \cite{lukiz33}, \cite{lukiz34} for $T=0$) and look for nonperturbative nonlinear solutions (see e.g. \cite{lukiz35}, \cite{lukiz36}).

 \end{document}